\documentclass[reprint,amssymb,amsmath,aps,superscriptaddress,showpacs,pra,footinbib,longbibliography]{revtex4-1}
\pdfoutput=1
\usepackage{graphicx}
\usepackage{color}

\newcommand{\Dv}{\mathbf D}
\newcommand{\Gammav}{\mathbf{\Gamma}}
\newcommand{\Sv}{\mathbf S}
\newcommand{\kv}{\mathbf k}
\newcommand{\iso}{\rm iso}
\newcommand{\eff}{\rm eff}
\newcommand{\FM}{\rm FM}
\newcommand{\AFM}{\rm AFM}
\newcommand{\eps}{\varepsilon}

\begin{document}

\title{Magnetic anisotropy in the frustrated spin chain compound $\beta$-TeVO$_4$}

\author{F. Weickert}
\email{weickert.ph@gmail.com}
% \email{mjaime@lanl.gov}
\author{N. Harrison}
\affiliation{MPA-CMMS, Los Alamos National Laboratory, Los Alamos, New Mexico 87545, USA}

\author{B.L. Scott}
\affiliation{MPA-11, Los Alamos National Laboratory, Los Alamos, New Mexico 87545, USA}

\author{M. Jaime}
\affiliation{MPA-CMMS, Los Alamos National Laboratory, Los Alamos, New Mexico 87545, USA}

\author{A. Leitm\"ae}
\author{I. Heinmaa}
\author{R. Stern}
\affiliation{National Institute of Chemical Physics and Biophysics, 12618 Tallinn, Estonia}

\author{O. Janson}
\affiliation{National Institute of Chemical Physics and Biophysics, 12618 Tallinn, Estonia}
\affiliation{Max Planck Institute for Chemical Physics of Solids, 01187 Dresden, Germany}
\affiliation{Institute of Solid State Physics, TU Wien, 1040 Vienna, Austria}

\author{H. Berger}
\affiliation{Ecole Polytechnique F\'{e}d\'{e}rale de Lausanne, Lausanne CH-1015, Switzerland}

\author{H. Rosner}
\affiliation{Max Planck Institute for Chemical Physics of Solids, 01187 Dresden, Germany}

\author{A.~A. Tsirlin}
\email{altsirlin@gmail.com}
\affiliation{National Institute of Chemical Physics and Biophysics, 12618 Tallinn, Estonia}
\affiliation{Max Planck Institute for Chemical Physics of Solids, 01187 Dresden, Germany}
\affiliation{Experimental Physics VI, Center for Electronic Correlations and Magnetism, Institute of Physics, University of Augsburg, 86135 Augsburg, Germany}

\begin{abstract}
Isotropic and anisotropic magnetic behavior of the frustrated spin chain compound $\beta$-TeVO$_4$ is reported. Three magnetic transitions observed in zero magnetic field are tracked in fields applied along different crystallographic directions using magnetization, heat capacity, and magnetostriction measurements. Qualitatively different temperature-field diagrams are obtained below 10\,T for the field applied along $a$ or $b$ and along $c$, respectively. In contrast, a nearly isotropic high-field phase emerges above 18\,T and persists up to the saturation that occurs around 22.5\,T. Upon cooling in low fields, the transitions at $T_{\rm N1}$ and $T_{\rm N2}$ toward the spin-density-wave and stripe phases are of the second order, whereas the transition at $T_{\rm N3}$ toward the helical state is of the first order and entails a lattice component. Our microscopic analysis identifies frustrated $J_1-J_2$ spin chains with a sizable antiferromagnetic interchain coupling in the $bc$ plane and ferromagnetic couplings along the $a$ direction. The competition between these ferromagnetic interchain couplings and the helical order within the chain underlies the incommensurate order along the $a$-direction, as observed experimentally. While a helical state is triggered by the competition between $J_1$ and $J_2$ within the chain, the plane of the helix is not uniquely defined because of competing magnetic anisotropies. Using high-resolution synchrotron diffraction and $^{125}$Te nuclear magnetic resonance, we also demonstrate that the crystal structure of $\beta$-TeVO$_4$ does not change down to 10\,K, and the orbital state of V$^{4+}$ is preserved.
\end{abstract}

\maketitle

\section{Introduction}
Frustrated (zigzag) spin-$\frac12$ chains with competing nearest-neighbor ferromagnetic ($J_1$) and next-nearest-neighbor antiferromagnetic ($J_2$) couplings reveal rich physics at low temperatures and in applied magnetic fields. When the chains are coupled in three dimensions, helical order arises in zero field for \mbox{$J_2/|J_1|>\frac14$~\cite{zinke2009,furukawa2010}.} While the helical order itself gives rise to a very unusual phenomenon of magnetic-field-induced ferroelectricity~\cite{park2007,schrettle2008,zhao2012}, further interesting effects occur when stronger magnetic fields break this order down. LiCuVO$_4$, one of the best studied frustrated-chain materials~\cite{[{For a brief review, see: }][]starykh2015}, undergoes a first-order transition around 8.5\,T from the helically-ordered phase toward a spin-density-wave (SDW) phase, where magnetic moments align with the field, and the length of the moment is modulated~\cite{buettgen2010,masuda2011}. Detailed nature of this phase is, however, debated~\cite{mourigal2012}, along with the putative nematic phase appearing around 40\,T right before saturation~\cite{svistov2011,buettgen2014}. Additionally, different types of multipolar order are expected for the $J_1-J_2$ chains in the applied magnetic field~\cite{hikihara2008,sudan2009,nishimoto2015}.

\begin{figure*}
\includegraphics{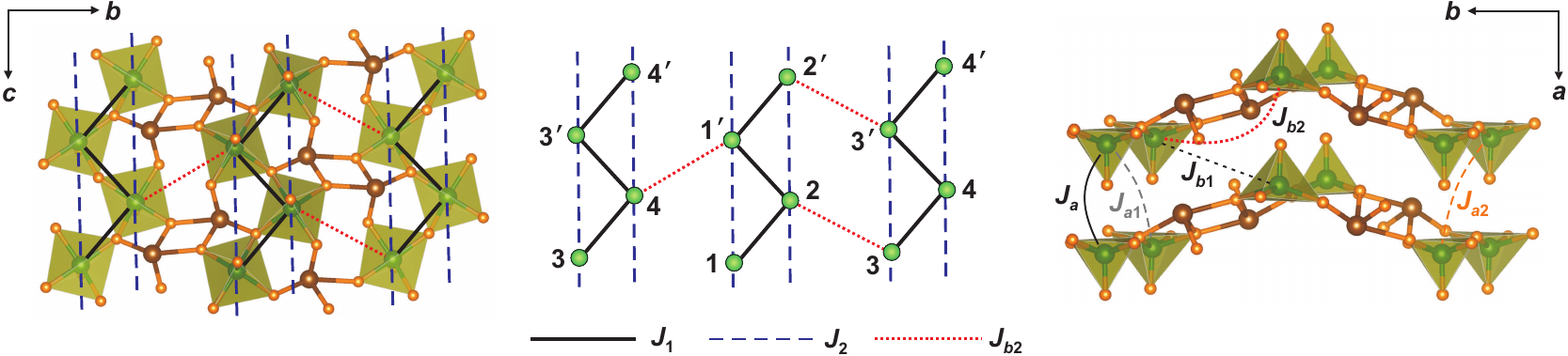}
\caption{\label{fig:structure}
(Color online) Crystal structure and magnetic model of $\beta$-TeVO$_4$. Left and middle panels: structural and magnetic layers composed of the $J_1-J_2$ frustrated spin chains. Right panel: interlayer couplings.
}
\end{figure*}
$\beta$-TeVO$_4$ is a candidate frustrated-chain material with spin-$\frac12$ (Fig.~\ref{fig:structure}). Its magnetic behavior was initially described within the model of a uniform spin-$\frac12$ chain~\cite{savina2011}, although the presence of three low-temperature transitions at $T_{\rm N1}\simeq 4.7$\,K, $T_{\rm N2}\simeq 3.3$\,K, and $T_{\rm N3}\simeq 2.3$\,K in an applied magnetic field as weak as 0.02\,T clearly indicates a more complex interaction topology. Recently, Sa\'ul and Radtke~\cite{saul2014} performed a microscopic analysis of isotropic exchange couplings and concluded that $\beta$-TeVO$_4$ is a good realization of the $J_1-J_2$ chain model with ferromagnetic (FM) $J_1$ and antiferromagnetic (AFM) $J_2$. Subsequently, magnetic susceptibility of $\beta$-TeVO$_4$ was re-analyzed in the framework of the $J_1-J_2$ model~\cite{savina2015a}. Gnezdilov \textit{et al.}~\cite{gnezdilov2012} reported non-monotonic evolution of phonon frequencies and speculated on possible structural changes around 150\,K and even on a change in the orbital state of V$^{4+}$ at low temperatures. Finally, Pregelj \textit{et al.}~\cite{pregelj2015} performed detailed neutron-scattering experiments in zero field and observed a helical magnetic structure with the propagation vector $\kv=(-0.208,0,0.423)$ below $T_{\rm N3}$. Between $T_{\rm N3}$ and $T_{\rm N2}$, $\beta$-TeVO$_4$ reveals an enigmatic stripe-like spin texture, whereas between $T_{\rm N2}$ and $T_{\rm N1}$ a spin-density-wave (SDW) phase akin to the field-induced SDW phase in LiCuVO$_4$~\cite{buettgen2010,masuda2011} has been proposed.

The crystal structure of $\beta$-TeVO$_4$~\cite{meunier1973} features chains of VO$_5$ polyhedra (Fig.~\ref{fig:structure}). These chains are directed along the crystallographic $c$-axis and linked via asymmetric TeO$_4$ units. The overall structure is centrosymmetric (space group $P2_1/c$), but inversion centers are located between the chains, so that Dzyaloshinsky-Moriya (DM) anisotropy terms are allowed for both $J_1$ and $J_2$, which is different from any of the frustrated-chain compounds previously reported. Magnetic anisotropy can have strong effect on field-induced phase transitions and on new phases induced by the magnetic field. For example, linarite PbCu(OH)$_2$SO$_4$, which symmetry is lower than in LiCuVO$_4$~\cite{wolter2012}, shows very complex and still poorly understood temperature-field phase diagrams~\cite{schaepers2013} that are weakly reminiscent of those for LiCuVO$_4$. 

In the following, we report a combined experimental and microscopic study of $\beta$-TeVO$_4$ and address several pending questions concerning this interesting material. In Sec.~\ref{sec:diagram}, we report temperature-field phase diagrams for different directions of the applied magnetic field and thus probe magnetic anisotropy in $\beta$-TeVO$_4$ experimentally. We show that $\beta$-TeVO$_4$ reveals very unusual phase diagrams with the first-order transition toward the helically ordered phase and second-order transitions between other phases. In Secs.~\ref{sec:structure} and~\ref{sec:nmr}, we discuss possible structural changes happening in the paramagnetic state and conclude that the overall symmetry of the structure as well as the orbital state of V$^{4+}$ are essentially unchanged down to at least $T_{\rm N1}$. Finally, in Sec.~\ref{sec:microscopic} we derive the microscopic spin Hamiltonian of $\beta$-TeVO$_4$, and briefly discuss its implications. Our results are summarized in Sec.~\ref{sec:summary}.

\section{Methods}
A slab-shaped single crystal of approximate dimensions $5\times 3\times 1$\,mm$^3$ was oriented using x-ray scattering, and found to have the longest dimension along the crystallographic $c$-axis and the shortest along the $a$-axis, while the intermediate dimension is parallel to the $b$-axis. It was experimentally found that while the $cb$-plane cleaves easily, exposing the chain-like underlying structure, neither the $ac$- nor the $ab$-plane do so.

Thermal expansion at constant magnetic fields and magnetostriction in pulsed magnetic field measurements were accomplished using an optical fiber Bragg grating (FBG) technique described before~\cite{jaime2012,daou2010}. Here, light reflected at the Bragg wavelength $\lambda_B$ by a grating inscribed in a telecom 125 $\mu$m diameter optical fiber is recorded with a spectrometer furbished with fast InGaAs line-array camera working at 46\,kHz~\cite{daou2010} and used to follow the sample dilation as the temperature and/or external magnetic field are changed. For these experiments the fiber was attached to the sample $ab$-plane when studying the strain along the $a$ and $b$ crystallographic direction, and to the $ac$-plane when studying the strain along the $c$ direction. 

The magnetization in pulsed magnetic fields to 60\,T at constant temperatures was obtained with a sample-extraction magnetometer working to $^3$He temperatures, and calibrated with measurements in a Quantum Design\textsuperscript{\textregistered} PPMS system to 14\,T. Specific heat measurements at constant magnetic fields were completed in the same PPMS system.

High-resolution x-ray diffraction (XRD) data were collected on the ID31 beamline of the European Synchrotron Radiation Facility (ESRF, Grenoble) using the wavelength of 0.4\,\r A. A small crystal of $\beta$-TeVO$_4$ was crushed, ground into fine powder and placed into a thin-wall borosilicate glass capillary that was spun during the measurement. The signal was collected by eight Si(111) analyzer crystals. Structure refinements were performed in the \texttt{JANA2006} program~\cite{jana2006}, and the resulting crystal structure was visualized using \texttt{VESTA}~\cite{vesta}.

All $^{125}$ Te NMR spectra were recorded on Bruker AVANCE-II NMR spectrometer at 14.1\,T magnetic field using home-built probe with a single-axis goniometer and He-flow cryostat from JANIS Research Inc. The temperature was monitored and regulated by a LakeShore-332 temperature controller. Each data point was obtained by recording the signal with a spin-echo sequence. The magnetic shift scale was referenced to the $^{125}$Te resonance frequency of Me$_2$Te 189.349\,MHz. 

The microscopic analysis of $\beta$-TeVO$_4$ is based on density-functional-theory (DFT) band-structure calculations performed in the \texttt{FPLO} code~\cite{fplo}. Experimental crystal structure obtained from low-temperature XRD has been used, and either the local-density approximation (LDA)~\cite{pw92} or generalized-gradient-approximation (GGA)~\cite{pbe96} exchange-correlation potentials were chosen. Isotropic exchange couplings were obtained from two complementary procedures, the LDA-based model analysis and the DFT+$U$ supercell calculations, as further explained in Sec.~\ref{sec:microscopic}. For the DFT+$U$ calculations, we used supercells doubled along either $a$ or $c$ directions. 

\begin{figure}
\includegraphics[width=0.9\columnwidth]{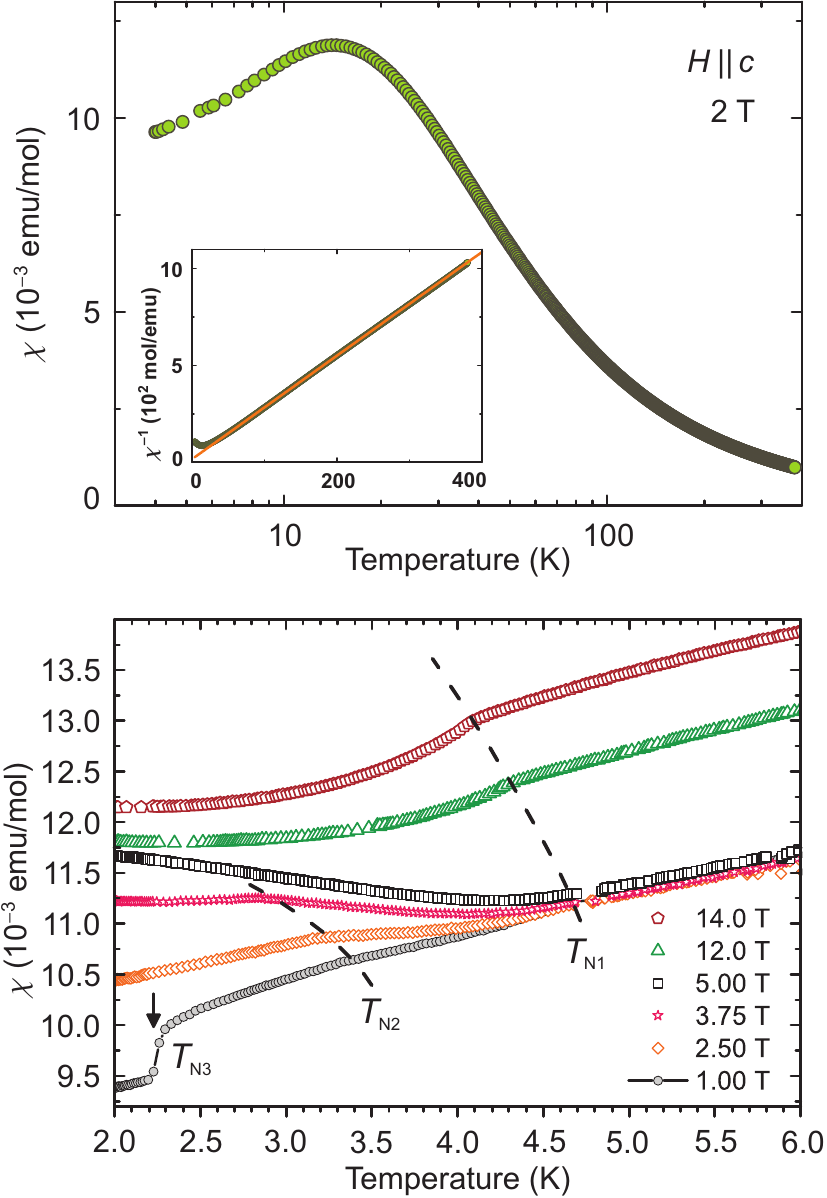}
\caption{(Color online) Top panel: Magnetic susceptibility vs temperature for $H=2$\,T in a wide temperature range. Inset: inverse magnetic susceptibility vs temperature showing Curie-Weiss behavior and fit. Bottom panel: magnetic susceptibility vs temperature in the low-temperature region at constant magnetic fields. Anomalies in $\chi(T)$ are indicated as $T_{\rm N1}$, $T_{\rm N2}$ and $T_{\rm N3}$, and dashed lines are guides to the eye.}
\label{fig:chi}
\end{figure}
\begin{figure}
\includegraphics[width=1\columnwidth]{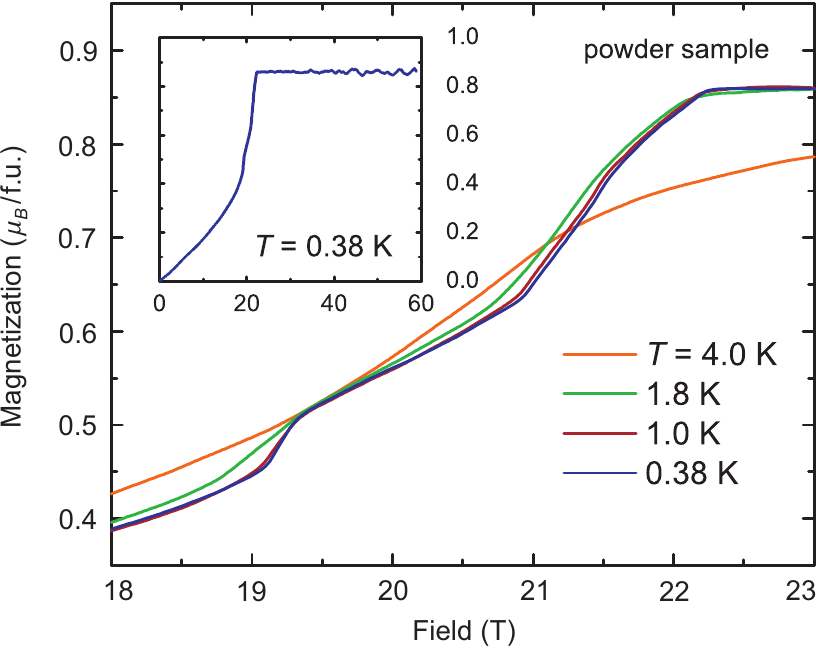}
\caption{(Color online) Magnetization vs magnetic field measured at temperatures between 0.38\,K and 4\,K in a powder sample. Note saturation at 0.86\,$\mu_B$/f.u. Inset: Magnetization vs magnetic field showing the full field range to 60\,T.}
\label{fig:mvsh}
\end{figure}

\begin{figure*}
\includegraphics{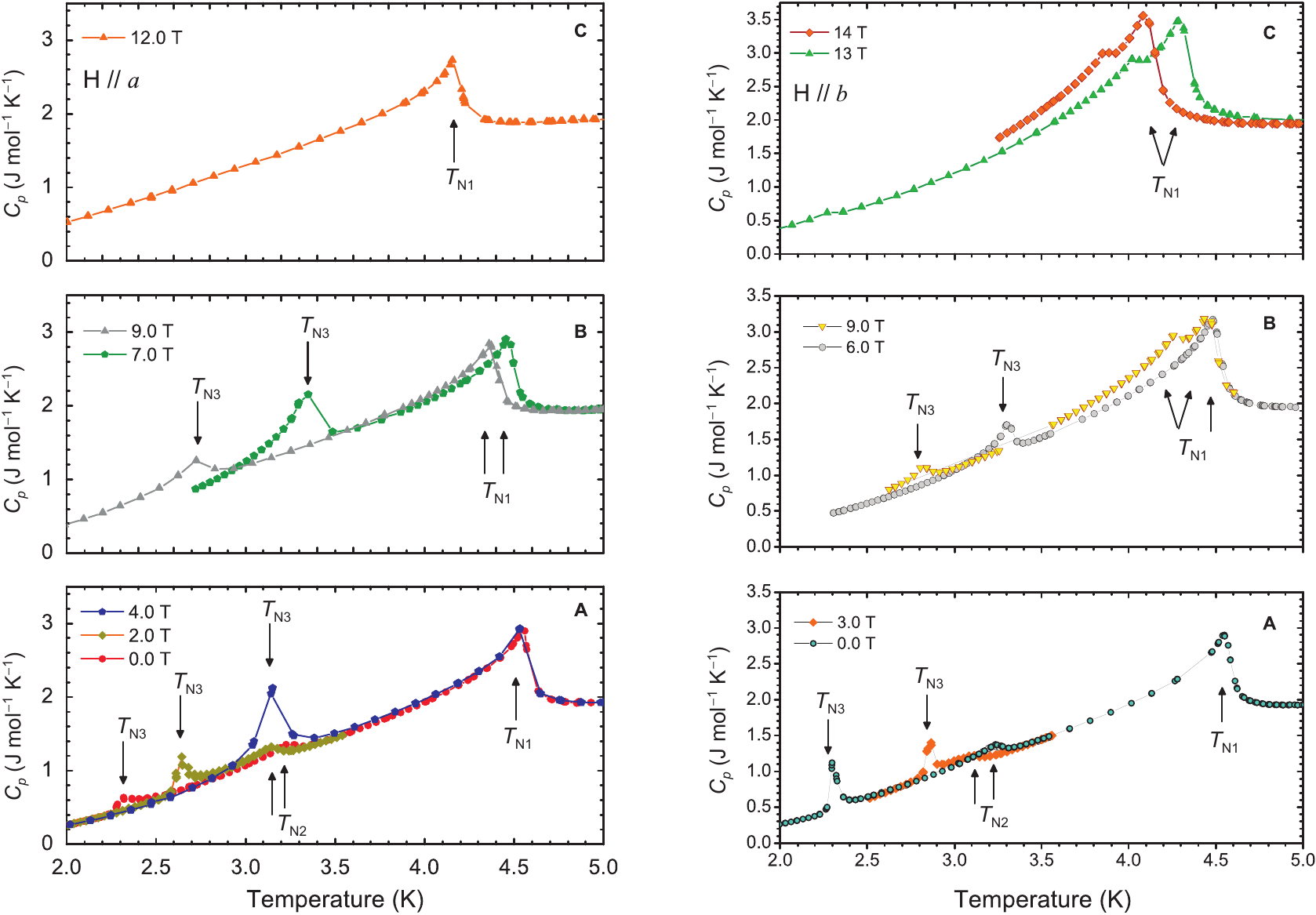}
\caption{\label{fig:heat1}\label{fig:heat2}
(Color online) (A) (B) (C) Specific heat vs temperature measured in the magnetic field applied along the $b$ (left) and $a$ (right) directions. Anomalies indicated with arrows are phase transitions.}
\end{figure*}
\begin{figure}[htb!]
\includegraphics[width=1\columnwidth]{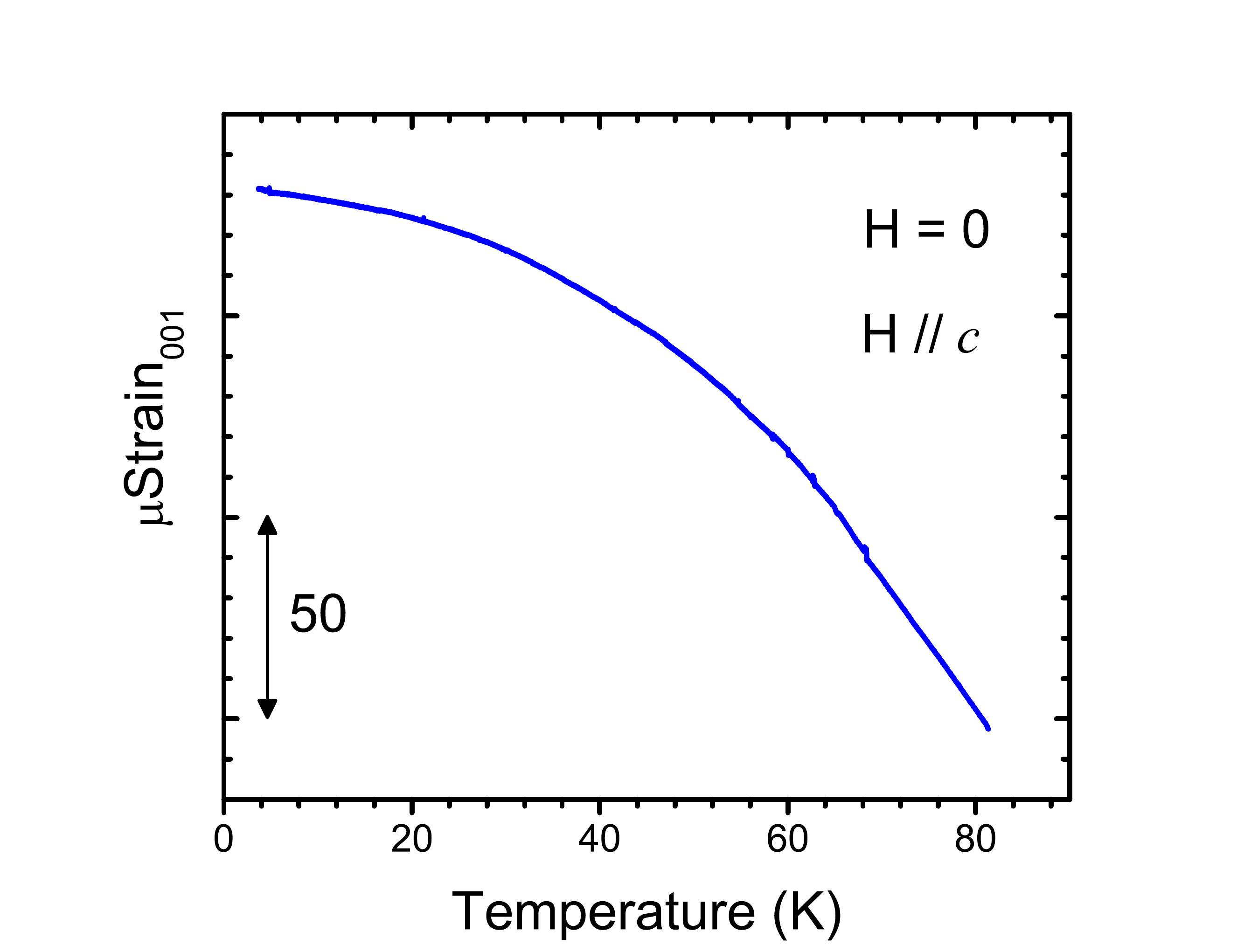}
\caption{(Color online) Contraction of the $c$-axis as a function of the increasing temperature in zero field, shown here to $T=80$\,K.}
\label{fig:expansion-c}
\end{figure}
Magnetic susceptibility for the $J_1-J_2$ chain was obtained by combining transfer-matrix renormalization group (TMRG)~\cite{wang1997} and exact-diagonalization (ED) simulations for the low- and high-temperature parts of the data, respectively. Exact diagonalization was performed for a finite lattice with $L=16$ sites and periodic boundary conditions. The use of TMRG helps to eliminate finite-size effects that manifest themselves at low temperatures. Field-dependent magnetization for the $J_1-J_2$ chain was calculated at zero temperature using density-matrix renormalization group (DMRG) method. The ED and DMRG simulations were performed in the \texttt{ALPS} simulations package~\cite{alps}.

\section{Results}
\subsection{Anisotropic magnetic properties and $(T,H)$ phase diagrams}
\label{sec:diagram}
The magnetic susceptibility ($\chi$) of a single-crystal sample was measured as a function of the temperature at constant magnetic fields between 0.5\,T and 14\,T. The results are displayed in Fig.~\ref{fig:chi}. At high temperatures, the magnetic susceptibility follows a Curie-Weiss dependence with a clear maximum centered at $T=14.5$\,K (Fig.~\ref{fig:chi}A), followed by several smaller anomalies as the temperature is reduced. A Curie-Weiss fit of the data in the $100-380$\,K range (Fig.~\ref{fig:chi}A, inset) gives a Curie constant $C=0.374$\,emu\,K\,mol$^{-1}$ and $\theta_{\rm CW}=-2.1$\,K indicating a nearly perfect balance of FM and AFM couplings, in agreement with earlier work \cite{gnezdilov2012}. The low-temperature anomalies are more clearly visible when the low temperature range is expanded (Fig.~\ref{fig:chi}B). Three phase transitions were identified as $T_{\rm N1}$, $T_{\rm N2}$ and $T_{\rm N3}$, and followed as the magnetic field was increased. While $T_{\rm N1}$ and $T_{\rm N2}$ share characteristics of second-order-like transitions, $T_{\rm N3}$ instead involves a rapid drop in $\chi$ suggesting a different, possibly first-order-like, process. Dashed lines indicate the evolution of them with applied field. 

\begin{figure*}
\includegraphics[width=1.8\columnwidth]{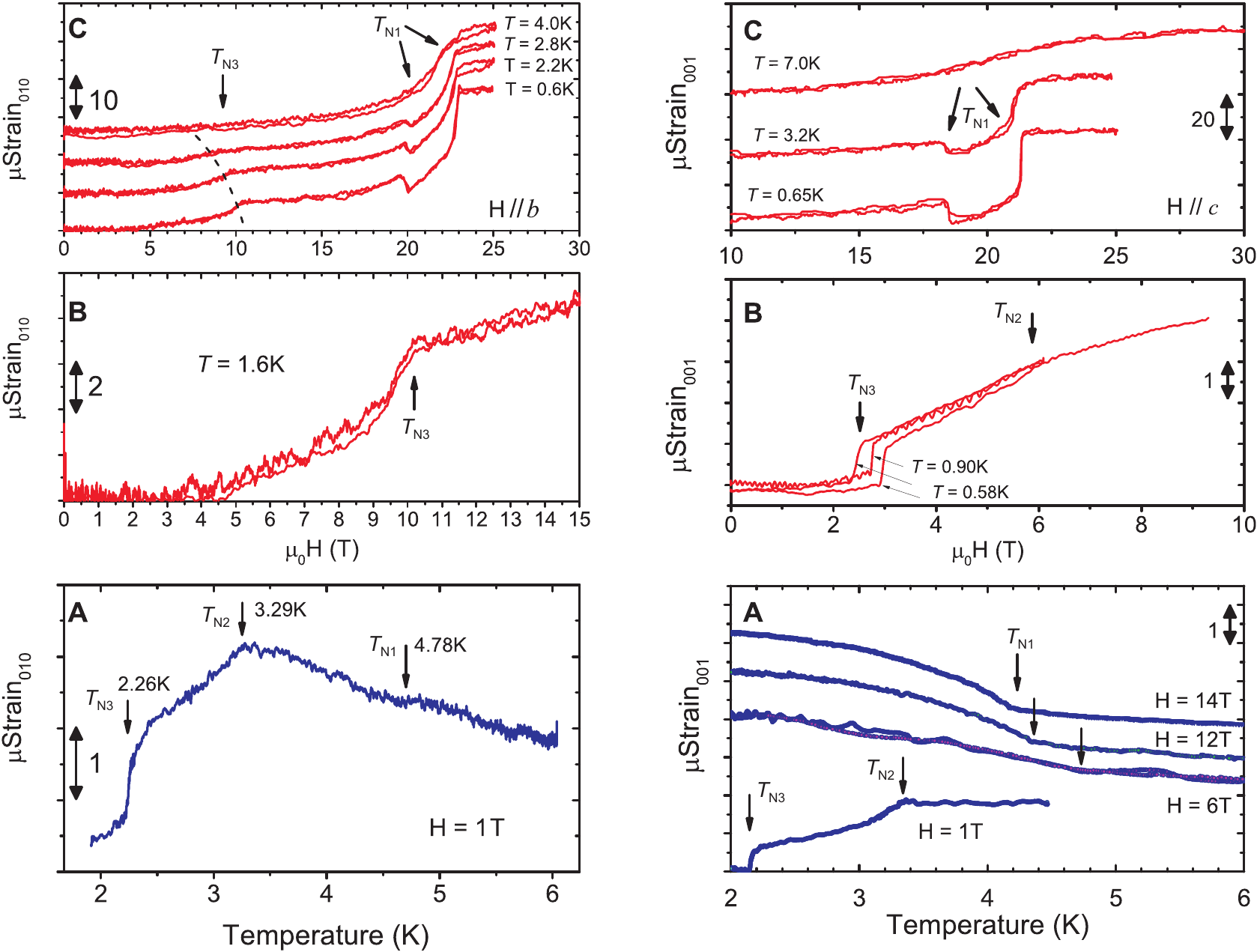}
\caption{(Color online) Magnetostriction measured in fields applied along the $b$ (left) and $c$ (right) directions. The strain is measured along the field. (A) Evolution of the sample length vs temperature at constant magnetic field $H=1$\,T, showing anomalies at transition temperatures indicated by arrows; (B) and (C) Magnetostriction in pulsed magnetic fields showing anomalies at fields indicated by arrows.}
\label{fig:expansion-b}
\label{fig:expansion-c2}
\end{figure*}
\begin{figure}
\includegraphics[width=1\columnwidth]{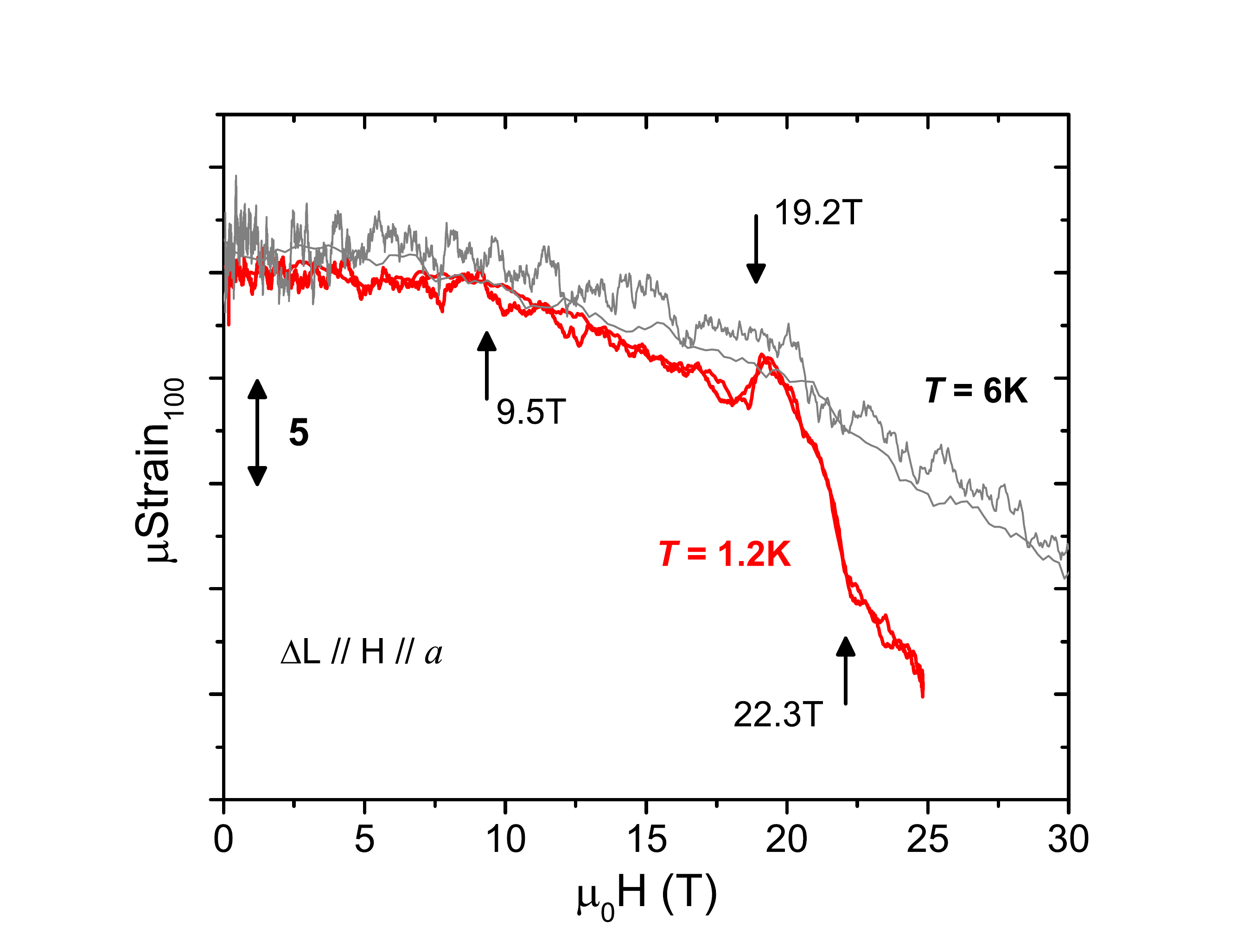}
\caption{(Color online) Magnetostriction measured along the $a$ direction in pulsed magnetic fields. Anomalies are indicated by arrows. As seen here, contrary to $b$- an $c$-axis, the $a$-axis length contracts with applied fields. The lack of saturation at fields $H>22$\,T could be an artifact caused by magnetic torque on the sample.  }
\label{fig:expansion-a}
\end{figure}
\begin{figure*}
\includegraphics{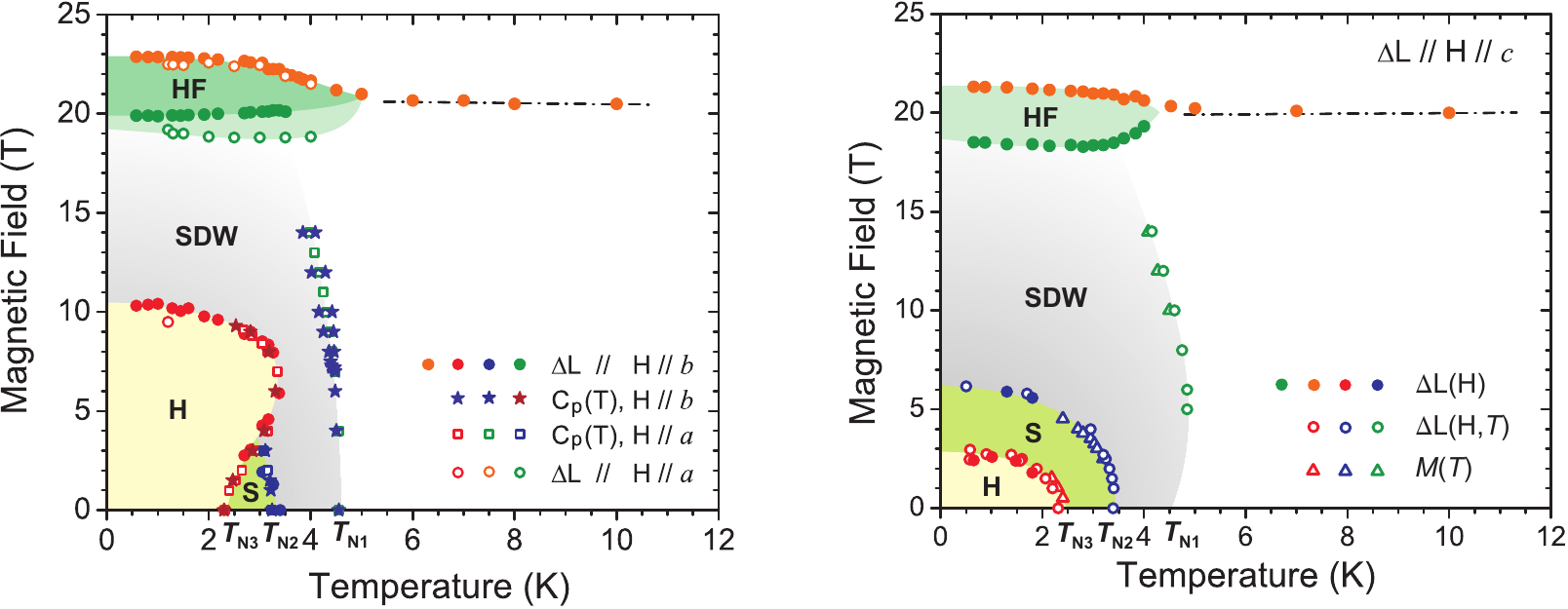}
\caption{(Color online) Field-Temperature phase diagram for $\beta$-TeVO$_4$ when the magnetic field is applied along the $a$ and $b$ axes (left) and along the $c$ axis (right). The phases are labeled as follows: H (helical order), S (long-period stripe phase), SDW (spin-density wave), and HF (high-field phase of unknown origin, possibly a nematic phase). }
\label{fig:diagram-a}\label{fig:diagram-c}
\end{figure*}
The magnetization vs field $M(H)$ measured in a powder sample, displayed in Figure~\ref{fig:mvsh}, shows $M(H)$ increasing gradually with field. Two small anomalies at 2.5\,T and 5.8\,T, not visible to the naked eye, are seen at $T=0.38$\,K (Fig.~\ref{fig:mvsh}, inset) and two large steps at 19 and 21\,T. These anomalies, likely associated to the suppression of low-temperature magnetic order, wash out and eventually vanish as the temperature is increased to 4\,K. A saturation magnetization of 0.86\,$\mu_{B}$/f.u. persists to the maximum field of 60\,T. Note that the saturation magnetization is somewhat below 1.0\,$\mu_B$ expected for a spin-$\frac12$ system. This might be due to ambiguities in scaling the pulsed-field data against the data collected in static fields.

The specific heat of a 6.2\,mg slab-shaped sample was measured in magnetic fields applied along the crystallographic $a$- and $b$-axes. Representative data is displayed in Fig.~\ref{fig:heat1} for different field directions. Particularly interesting are the $H\| b$-axis data: here, the transition $T_{\rm N3}=2.3$\,K in zero field looks different from the other $\lambda$-like transitions, much as in the case of $M(T)$ data, reinforcing the support for a different type of transition. Also, along this direction the transition $T_{\rm N1}$ is observed to split for magnetic fields $H\geq 9$\,T.

The nature of low-temperature phase transitions in $\beta$-TeVO$_4$ can be tested with probes that are sensitive to the crystal lattice, such as thermal expansion and magneto-strain $\Delta L/L(H)$, to gather information on the relevance and strength of spin-lattice correlations. Figure~\ref{fig:expansion-c} shows a negative thermal expansion along the $c$ direction ($\mu$Strain $\equiv$ 10$^{-6} \Delta L/L$), $i.e.$ the sample expands as the temperature drops. Similar behavior was observed for the $b$ direction. Figure~\ref{fig:expansion-b}A displays the thermal expansion at constant magnetic field $H=1$\,T, in the low temperature range where the transitions at $T_{\rm N1}$, $T_{\rm N2}$, and $T_{\rm N3}$ are clearly visible. While $T_{\rm N1}$ and $T_{\rm N2}$ look like conventional $\lambda$-like second-order transitions, $T_{\rm N3}$, again, looks different. The sharp drop in lattice parameter on cooling resembles a first-order like transition. It is often the case that phase transitions displaying important lattice involvement become first order and develop hysteresis. In this case, however, hysteresis was not significant. We must point out, however, that our optical fiber technique is not the most appropriate for detection of thermal hysteresis as the fiber (the only contact between sample and sample holder) is a very poor thermal conductor and some thermal lag between sample and thermometer is always present. Figures~\ref{fig:expansion-b}B and \ref{fig:expansion-b}C display the magnetostriction obtained in a pulsed magnet to $H=30$\,T. The transition $T_{\rm N3}$ is visible as an increase in the lattice parameter with field, while $T_{\rm N1}$ appears split first as a decrease of the lattice parameter at 20\,T then as a larger increase at 23\,T.  

Figure~\ref{fig:expansion-c2}A (right) shows the thermal expansion measured along the chains direction $c$ at constant magnetic fields in the low temperature region. $T_{\rm N1}$, not visible in zero field, becomes clear as the magnetic field is increased to 14T. $T_{\rm N2}$ and $T_{\rm N3}$, on the other hand, look very similar to the $H\| b$ case. Figures~\ref{fig:expansion-c2}B and \ref{fig:expansion-c2}C show the magnetostriction measured in pulsed magnetic fields, where the transitions $T_{\rm N3}$, $T_{\rm N2}$ and $T_{\rm N1}$ are clearly identified. While these transitions maintain similar characteristics for $H\|b$ and $H\|c$, the critical fields are significantly different. Figure~\ref{fig:expansion-a} displays the magnetostriction in pulsed fields measured along the $a$-axis, which shows anomalies at approximately 19\,T and 22.5\,T. At these anomalies the $a$-axis behaves opposite to the $b$- and $c$-axes, pointing to a partial compensation of the expansion trend shown by those. The unit cell volume, hence, appears to expand as the sample magnetization is saturated by external magnetic fields. 

Critical temperatures and critical fields identified in the above discussed magnetization $M(T,H)$, specific heat $C_p(T)$, thermal expansion $\mu$Strain($T$), and magnetostriction $\mu$Strain($H$) were extracted and plotted in two phase diagram displayed in Figure~\ref{fig:diagram-a}. These plots reveal the previously unknown and intricate anisotropy, and the possibility of a tricritical point at $H=20$\,T, and $T\simeq 4$\,K. Using zero-field neutron data~\cite{pregelj2015}, we identify the phase below $T_{\rm N3}$ as the helical order (H), the phase between $T_{\rm N2}$ and $T_{\rm N3}$ as the long-period stripe order (S), and the phase between $T_{\rm N1}$ and $T_{\rm N2}$ as the SDW order. The low-temperature helical phase is thus relatively stable in magnetic fields applied along $a$ and $b$ and becomes fragile when the field is applied along $c$, where, on the other hand, a broader region of the stripe phase is observed. Our observations are consistent with the results of Ref.~\onlinecite{savina2015b}, where temperature-field phase diagrams for two different field directions ($a$ and $c$ not resolved) were reported based on magnetic susceptibility measurements and below 5\,T, only.

\subsection{Crystal structure}
\label{sec:structure}
\begin{table}
\caption{\label{tab:coordinates}
Atomic positions in $\beta$-TeVO$_4$ refined from high-resolution synchrotron powder XRD. $U_{\iso}$ are isotropic atomic displacements parameters (ADPs) given in 10$^{-2}$\r A$^{2}$. For each atom, the first line refers to the 10\,K data, and the second line refers to the room-temperature data. The ADPs of oxygen atoms were constrained during the refinement. Lattice parameters are $a=4.33989(2)$\,\r A, $b=13.4943(1)$\,\r A, $c=5.44460(3)$\,\r A, $\beta=91.6572(2)^{\circ}$ at 10\,K and $a=4.38126(2)$\,\r A, $b=13.5089(1)$\,\r A, $c=5.44201(3)$\,\r A, $\beta=91.6766(3)^{\circ}$ at room temperature. The space group is $P2_1/c$.
}
\begin{ruledtabular}
\begin{tabular}{ccccc}
   & $x/a$     & $y/b$     & $z/c$     & $U_{\iso}$ \\
Te & 0.0421(1) & 0.3911(1) & 0.6431(1) & 0.31(1) \\
   & 0.0392(1) & 0.3911(1) & 0.6430(1) & 0.99(2) \\
V  & 0.6783(3) & 0.1602(1) & 0.6611(2) & 0.36(3) \\
   & 0.6798(3) & 0.1611(1) & 0.6598(3) & 0.65(4) \\
O1 & 0.3095(8) & 0.1640(3) & 0.6679(6) & 0.18(5) \\
   & 0.296(1)  & 0.1643(4) & 0.6674(8) & 1.03(7) \\
O2 & 0.8310(8) & 0.0480(3) & 0.8633(7) & 0.18(5) \\
   & 0.840(1)  & 0.0509(4) & 0.8550(9) & 1.03(7) \\
O3 & 0.8129(8) & 0.2230(3) & 0.9794(6) & 0.18(5) \\
   & 0.816(1)  & 0.2244(4) & 0.9824(9) & 1.03(7) \\
O4 & 0.7501(8) & 0.0828(2) & 0.3713(6) & 0.18(5) \\
   & 0.759(1)  & 0.0837(3) & 0.3738(8) & 1.03(7) \\
\end{tabular}
\end{ruledtabular}
\end{table}
The intricate temperature-field phase diagrams presented above can not be understood without solid knowledge of the underlying crystal structure. According to Ref.~\cite{gnezdilov2012}, structural changes in $\beta$-TeVO$_4$ may happen around 150\,K in the paramagnetic regime well above any magnetic transitions. To investigate this possibility, we performed high-resolution synchrotron XRD measurements at room temperature and at 10\,K. No reflections violating the $P2_1/c$ symmetry could be seen in either of the patterns. Structure refinement revealed only weak temperature-induced structural changes. Most importantly, atomic displacements parameters decrease upon cooling, as expected in a well-ordered crystal structure. Therefore, we conclude that no drastic structural changes occur in $\beta$-TeVO$_4$ down to at least 10\,K, and the orbital state of V$^{4+}$ is robust. Our thermodynamic and magnetostriction measurements presented in Sec.~\ref{sec:diagram} further rule out any drastic structural changes at $T_{\rm N1}$ and $T_{\rm N2}$, whereas the transition at $T_{\rm N3}$ should be coupled to the lattice. It is, however, well below the temperature range accessible for synchrotron XRD.

Upon cooling from room temperature to 10\,K, the unit cell volume decreases by about 1\,\%. Remarkably, this change is mostly related to the contraction of the $a$ parameter, whereas the $c$ parameter even increases by 0.05\,\% in agreement with strain measurements in zero field (Fig.~\ref{fig:expansion-c}). This strongly anisotropic thermal expansion can be traced back to peculiarities of the crystal structure. The structural chains running along the $c$ direction are linked in the $bc$ plane via TeO$_4$ pyramids (Fig.~\ref{fig:structure}). The interlayer bonding is achieved via longer and thus weaker Te--O1 bonds of 2.95\,\r A, hence the $a$ direction is most prone to expansion upon heating. The weak bonding along $a$ is also consistent with the preferential cleaving of the $\beta$-TeVO$_4$ crystals perpendicular to the $a$ direction.

\subsection{$^{125}$Te NMR}
\label{sec:nmr}
XRD probes long-range crystal structure in the bulk and may be less sensitive to structural changes that occur locally. Therefore, we also studied temperature evolution of $\beta$-TeVO$_4$ above its magnetic transitions using $^{125}$Te NMR. The resonance frequency of the spin-1/2 $^{125}$Te nucleus is determined by chemical shift interaction or, in magnetic materials, by the magnetic hyperfine shift, the Knight shift, interaction. In solids as a rule, both interactions are anisotropic and described by a second-rank tensor. There are two pairs of tellurium Te$^{4+}$ ions in the crystallographic unit cell (see Fig.~\ref{fig:structure}, left). For an arbitrary direction of the magnetic field, the two neighboring Te ions of the same pair make equal projections, whereas the Te ions of the other pair have a different projection. Thus, one expects two $^{125}$Te resonance lines in the spectrum of a single crystal at an arbitrary orientation of the magnetic field. Indeed, experiment shows two lines, which we denote as site 1 and site 2. 

\begin{figure}
\includegraphics[width=0.9\columnwidth]{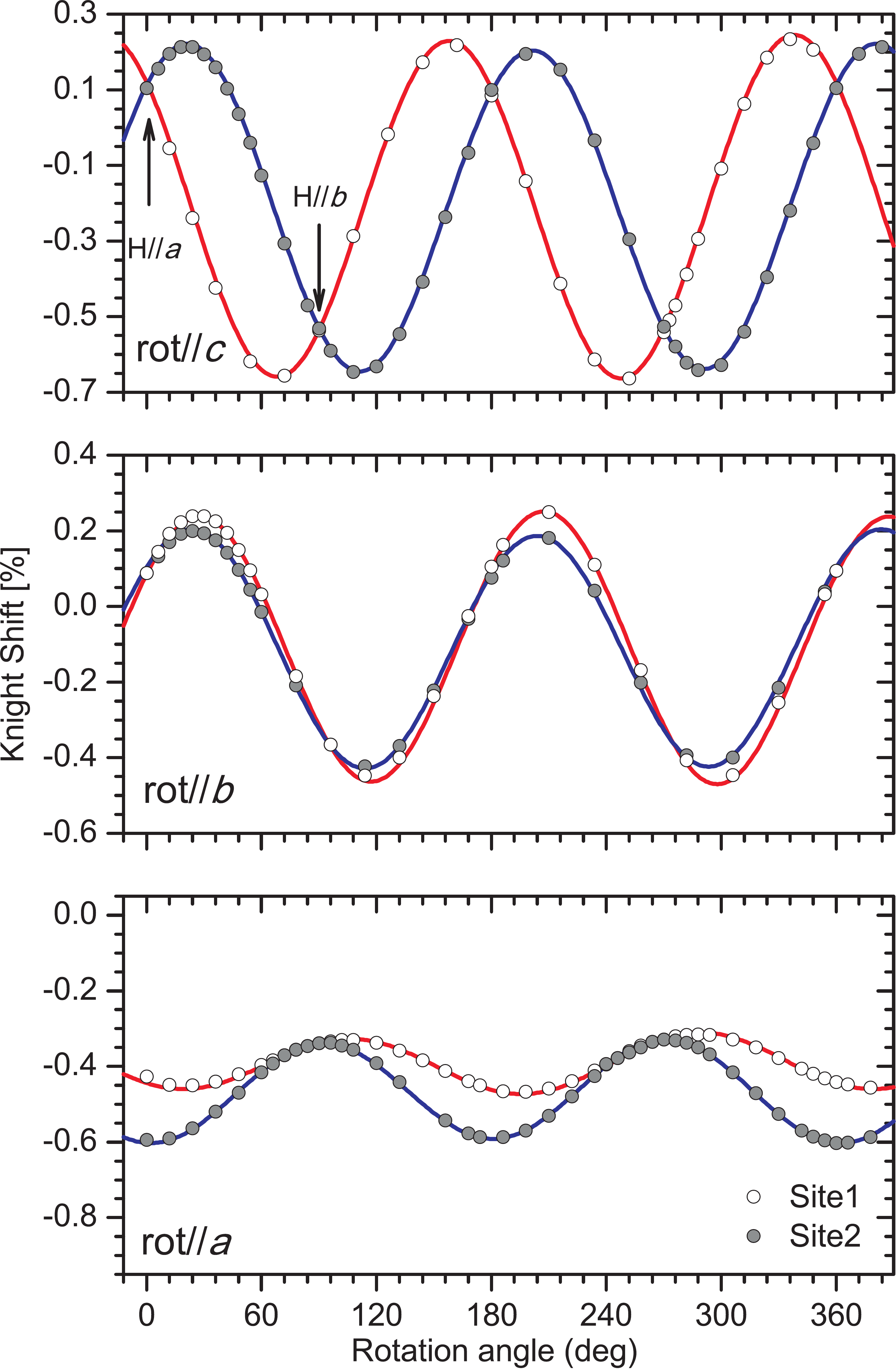}
\caption{\label{fig:nmr-rotation}
(Color online) $^{125}$Te Knight shift versus goniometer rotation angle measured on the $\beta$-TeVO$_4$ single crystal at 285\,K. The single crystal is rotated about the $a$-, $b$- and $c$-axes, as noted in the figure. The circles denote experimental frequencies, the full lines show the calculated angular dependence. The arrows denote orientations, where the temperature dependencies were measured.
}
\end{figure}
To determine the magnetic shift tensor, we performed three rotations of the single crystal around (approximately) $a$-, $b$-, and $c$-axes. The rotation patterns are given in Figure~\ref{fig:nmr-rotation}. Following the standard procedure~\cite{mehring1976}, one has to perform three subsequent transformations of the nuclear spin Hamiltonian: from the principal axis system (PAS) to the crystal frame, then to the goniometer frame, and eventually to the lab frame, and then the eigenvalues can be found. This way, we obtained the Knight shift tensor $K$ with the principal components $K_{xx}=-0.67$\%, $K_{yy}=-0.40$\%, and $K_{zz}=+0.32$\% (and the resulting isotropic value of $K_{\rm iso}=-0.25$\%) equal for both Te$^{4+}$ sites. The Euler angles $(\alpha;\beta;\gamma)$ for transforming the PAS of the $K$-tensor to the crystal frame are $(106; 69; 110)$ degrees for site 1 and  $(-106; -69; -110)$ degrees for site 2.

\begin{figure}
\includegraphics[width=1\columnwidth]{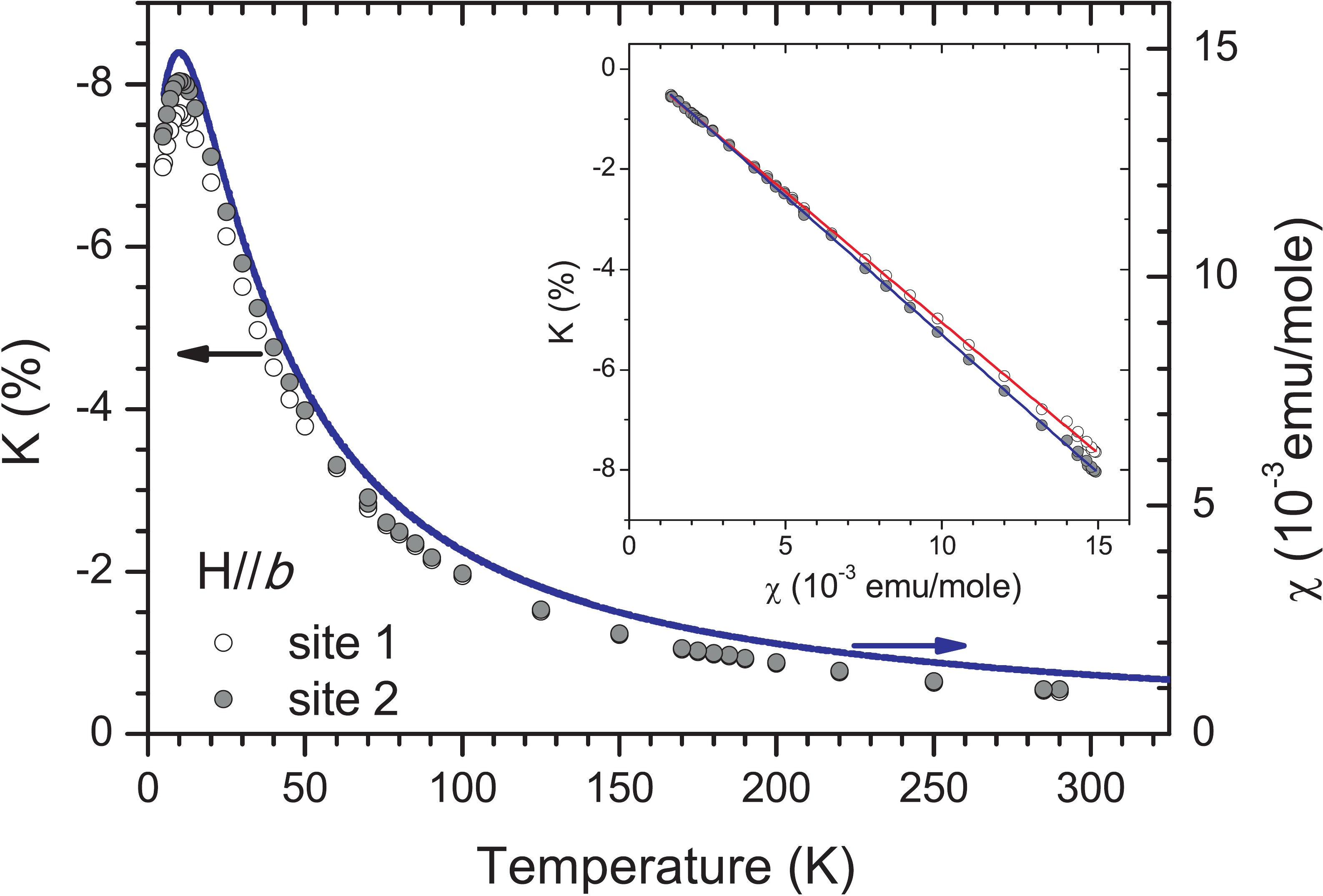}
\caption{\label{fig:nmr-shift}
(Color online) Temperature dependence of the Knight shift $K$ (full and empty circles, left scale) in the orientation $H\|b$ compared to the magnetic susceptibility curve for $H\|b$ (right scale, solid line) measured at 14\,T. The two sites have slightly different orientations of the $K$-tensor. The inset shows the perfectly linear $K$ vs. $\chi$ plot, where the slope gives the hyperfine coupling constant (see text). 
}
\end{figure}
At lower temperatures, the NMR lines broaden, although no abrupt changes are observed around 150\,K, where Gnezdilov~\textit{et al.}~\cite{gnezdilov2012} expected a structural phase transition. The Knight shift ($K$) follows bulk magnetic susceptibility ($\chi$), as shown in Fig.~\ref{fig:nmr-shift}. From the slope of the Clogston-Jaccarino plot~\cite{clogston1961} $K$ vs. $\chi$ (inset of Fig.~\ref{fig:nmr-shift}), we determine the hyperfine coupling constant as $H_{\rm hf}=2N_A\mu_B\Delta K/\Delta\chi$, where $N_A$ is Avogadro's number, and $\mu_B$ is the Bohr magneton. For the two slightly differently oriented $K$-tensors of Te, we obtain $H_{\rm hf}=-55$\,kOe/$\mu_B$ and $H_{\rm hf}=-58$\,kOe/$\mu_B$. 

\begin{figure}
\includegraphics[width=0.9\columnwidth]{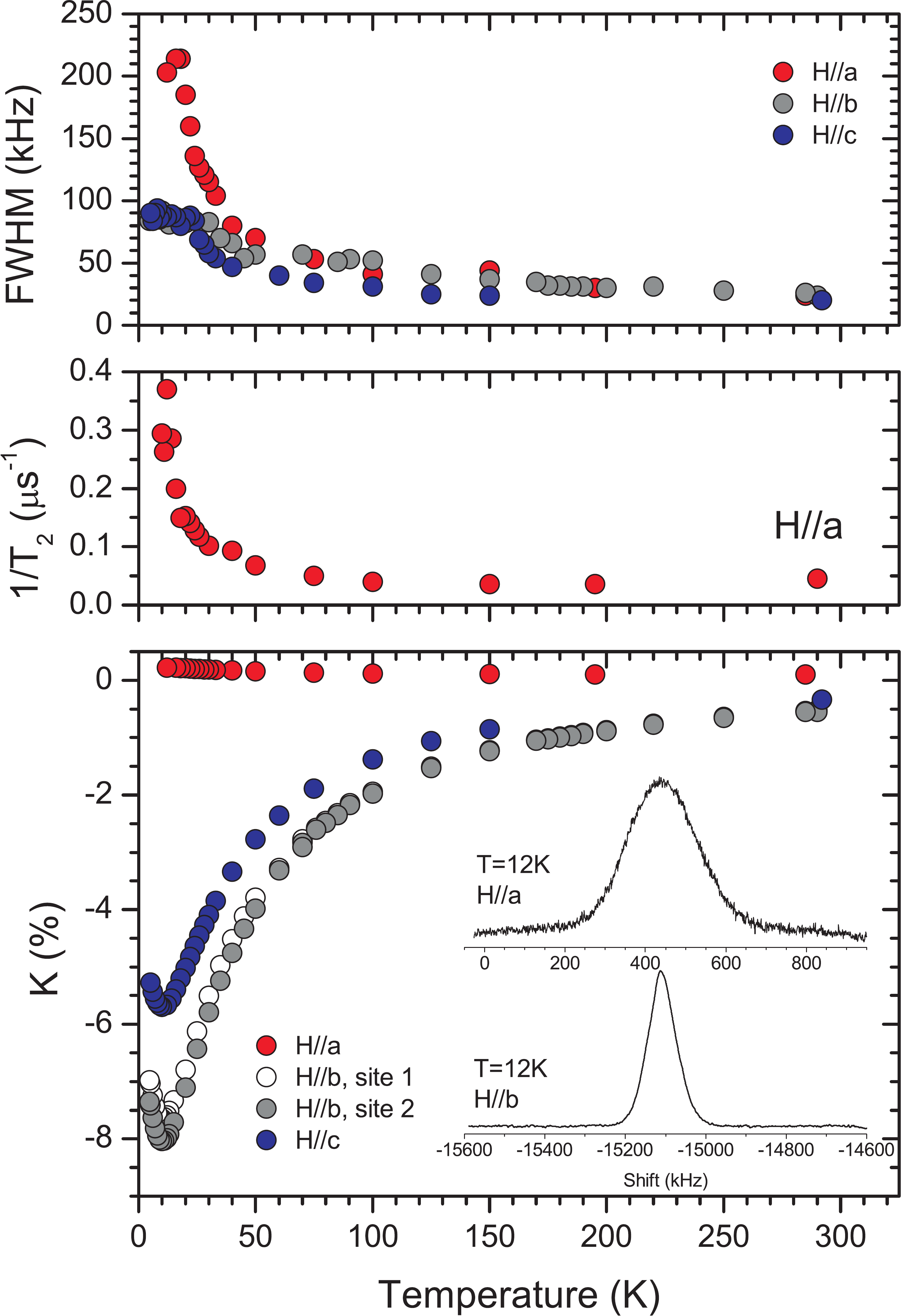}
\caption{\label{fig:nmr-relaxation}
(Color online) Temperature dependence of the NMR linewidth (top panel), of the transverse relaxation rate in the $H\|a$ orientation (middle) and the Knight shift curves in three different orientations of the crystal. The inset shows the NMR lines recorded for the $H\|a$ and $H\|b$ orientations at 12\,K.
}
\end{figure}

Compared to the $H\| b$ case, the Knight shift in the $H\| a$ orientation is small, and its temperature dependence is rather weak (see Figure~\ref{fig:nmr-relaxation}, bottom). On the other hand, at low temperatures the line for $H\| a$ becomes nearly three times broader than for the other orientations. The line broadening in the $H\| a$ direction is accompanied by an increase in the transverse relaxation rate $1/T_2$. Below 10\,K, $T_2$ became shorter than few $\mu$s, so that we could not record the line. For the two other directions, $T_2$ decreases by less than a factor of two, from 50\,$\mu$s at 290\,K to 30\,$\mu$s at 10\,K, from which down to $T_{\rm N1}$ it shortens again down to few $\mu$s. Rapid shortening of the $T_2$ refers typically to the zero-frequency fluctuation of the local field along the external field. The spin-lattice relaxation time $T_1$ is found to be about 50\,$\mu$s, almost independent of temperature and the crystal orientation.

\subsection{Microscopic analysis}
\label{sec:microscopic}
The complex magnetic behavior of $\beta$-TeVO$_4$ hinges upon its non-trivial crystal structure that gives rise to both isotropic and anisotropic magnetic couplings. Having established that the crystal structure remains unchanged down to $T_{\rm N1}$, we will now use it to parametrize a microscopic spin Hamiltonian:
\begin{equation}
 \hat H=\sum_{\langle ij\rangle}\left(J_{ij}\Sv_i\cdot\Sv_j+\mathbf D_{ij}[\Sv_i\times\Sv_j]+\Sv_i\cdot\Gammav_{ij}\cdot\Sv_j\right),
\label{eq:ham}\end{equation}
where the summation is over bonds of the spin lattice, $J_{ij}$ are isotropic exchange couplings, $\Dv_{ij}$ are DM vectors (antisymmetric part of the exchange anisotropy), and $\Gammav_{ij}$ tensors stand for the symmetric part of the anisotropy. We will evaluate the isotropic part of this Hamiltonian ($J_{ij}$) and rationalize several peculiarities of the low-temperature behavior. We will also analyze the anisotropic part qualitatively in order to underpin the difference between $\beta$-TeVO$_4$ and other $J_1-J_2$ frustrated-chain compounds.

\subsubsection{Isotropic couplings}
\begin{figure}
\includegraphics[width=0.9\columnwidth]{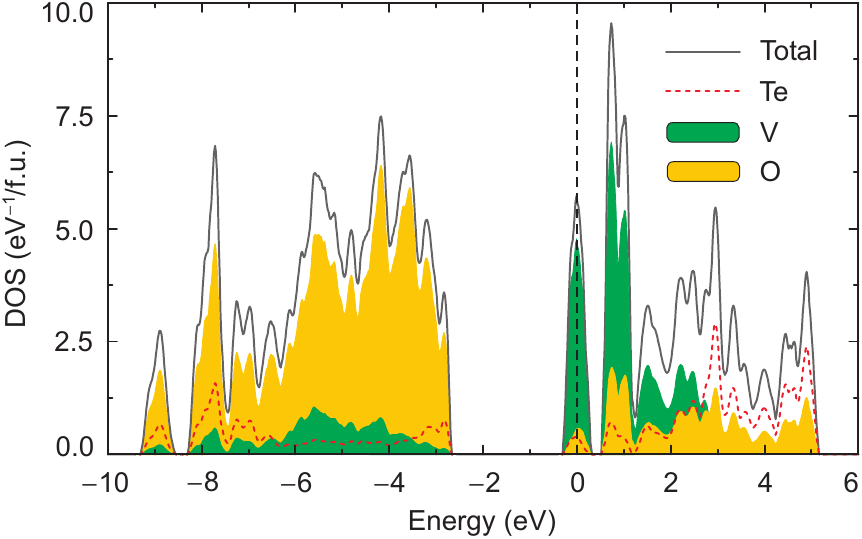}
\caption{\label{fig:dos}
(Color online) LDA density of states for $\beta$-TeVO$_4$. The Fermi level is at zero energy. Note the crystal-field splitting of V $3d$ states and their hybridization with Te $5p$ above 2.5\,eV.
}
\end{figure}

All calculations were performed for the experimental crystal structure of $\beta$-TeVO$_4$ determined at 10\,K (Table~\ref{tab:coordinates}). Band structure calculated on the LDA level is shown in Fig.~\ref{fig:dos}. Its apparent metallicity is related to the fact that LDA does not capture effects of strong electronic correlations pertinent to the $3d$ shell of V$^{4+}$~\footnote{Note that in DFT+$U$ we reproduce the insulating state with a band gap of about 3.0\,eV.}. We find predominantly V $3d$ states at the Fermi level. Crystal-field effects split the V $3d$ states into two narrow band complexes that can be ascribed~\footnote{Here, we used the coordinate frame with the $z$-axis directed along the short V--O bond in the axial position of the VO$_5$ square pyramids. The $x$ and $y$ axes are directed roughly toward the oxygen atoms in the basal plane of the pyramid, subject to a condition that $x$ and $y$ are perpendicular to $z$.} to the $d_{xy}$ and $d_{yz}+d_{xz}$ orbitals around 0\,eV and 1\,eV, respectively. Above 1.2\,eV, the spectrum is dominated by the $d_{x^2-y^2}+d_{3z^2-r^2}$ orbitals that strongly hybridize with Te $5p$-orbitals above 2.5\,eV. The $d_{xy}$-states have the lowest energy, in agreement with the crystal-field splitting expected for the 5-fold oxygen coordination in the square pyramid.

The LDA band structure was analyzed using a 12-band tight-binding model, where we included all $d_{xy}$, $d_{yz}$, and $d_{xz}$ bands lying between $-0.3$\,eV and $1.0$\,eV. The hopping parameters $t$ are then used to calculate individual exchange integrals according to the Kugel-Khomskii model~\cite{mazurenko2006,tsirlin2011}:
\begin{equation}
 J=\dfrac{4t_{xy\rightarrow xy}^2}{U_{\eff}}-\sum_{\alpha=yz,xz}\dfrac{4J_{\eff}t_{xy\rightarrow\alpha}^2}{(U_{\eff}+\Delta_{\alpha})(U_{\eff}-J_{\eff}+\Delta_{\alpha})},
\label{eq:kk}\end{equation}
where $\eps_{\alpha}$ are orbital energies, $\Delta_{\alpha}=\eps_{\alpha}-\eps_{xy}$ are crystal-field splittings, and $U_{\eff}=4$\,eV and $J_{\eff}=1$\,eV are the effective Coulomb repulsion and Hund's exchange in the V $3d$ shell~\cite{tsirlin2011b,tsirlin2011c}. The higher-lying $d$-orbitals ($\alpha=3z^2-r^2,x^2-y^2$) were excluded because of their strong mixing with the Te $5p$-states that renders the fitting procedure somewhat ambiguous. We checked, however, that different fits, where the higher-lying orbitals were included, produce similar results and do not influence any of the conclusions drawn below.

In Table~\ref{tab:exchanges}, we list both FM and AFM contributions to the exchange corresponding to the second and first terms of Eq.~\eqref{eq:kk}, respectively. Alternatively, we evaluated exchange couplings by calculating total energies of several collinear spin configurations using DFT+$U$. Here, we used the generalized gradient approximation (GGA)~\cite{pbe96}, the on-site Coulomb repulsion $U_d=4$\,eV, the on-site Hund's exchange $J_d=1$\,eV, and the atomic-limit version of the double-counting correction term. We have also checked that a different double-counting correction scheme or different values of $U_d$ lead to marginal changes in the exchange couplings without altering the resulting magnetic model.

\begin{table}
\caption{\label{tab:exchanges}
Interatomic distances $d_{\text{V--V}}$ (in~\r A) and isotropic exchange couplings $J_i$ (in~K) obtained from Eq.~\eqref{eq:kk} and from total-energy GGA+$U$ calculations. $J_i^{\AFM}$ and $J_i^{\FM}$ stand for the AFM and FM contributions to the exchange, according to the first and second terms of Eq.~\eqref{eq:kk}, respectively. For the notation of exchange couplings, see Fig.~\ref{fig:structure}.
}
\begin{ruledtabular}
\begin{tabular}{ccrrrr}
      & $d_{\text{V--V}}$ & $J_i^{\AFM}$ & $J_i^{\FM}$ & $J_i^{\rm Eq.~\eqref{eq:kk}}$ & $J_i^{\text{GGA}+U}$ \\
$J_1$ &     3.643         &     0.2      &  $-41.8$  & $-41.6$ & $-26.2$ \\
$J_2$ &     5.445         &    71.4      &   $-3.7$  &   67.7  &  24.6   \\
$J_a$ &     4.340         &     0.5      &   $-1.1$  &  $-0.6$ &  $-0.5$ \\
$J_{a1}$ &  5.603         &     0.0      &   $-1.8$  &  $-1.8$ &  $-2.2$ \\
$J_{a2}$ &  5.726         &     0.7      &   $-1.2$  &  $-0.5$ &  $-0.5$ \\
$J_{b1}$ &  4.902         &     5.3      &   $-4.3$  &    1.0  &   1.0   \\
$J_{b2}$ &  5.464         &    25.1      &   $-2.3$  &   22.8  &   7.3   \\
\end{tabular}
\end{ruledtabular}
\end{table}

Our results for the isotropic exchange couplings are in line with those from Ref.~\onlinecite{saul2014}. The two leading interactions are FM $J_1$ and AFM $J_2$ forming frustrated spin chains along the $c$ direction. The strongest interchain coupling $J_{b2}$ connects the chains in the $bc$ plane, within the structural layers. The coupling between the layers is facilitated by FM $J_a$, $J_{a1}$, and $J_{a2}$. Other couplings along the $a$ direction are below 0.2\,K (by absolute value), and the couplings in the $bc$ plane beyond $J_{b2}$ are 2.0\,K or less. Therefore, the magnetic model with $J_1$ and $J_2$ as intrachain couplings and $J_{b2}$, $J_a$, $J_{a1}$, and $J_{a2}$ as interchain couplings provides an exhaustive microscopic description of $\beta$-TeVO$_4$ on the isotropic (Heisenberg) level.

The microscopic origin of these couplings can be understood as follows. The coupling $J_1$ involves only one oxygen atom and corresponds to the V--O--V pathway with the bridging angle of $133.6^{\circ}$. While Goodenough-Kanamori-Anderson rules prescribe that such a coupling should be AFM, its AFM part is in fact negligible, and the FM part dominates. This is in line with the results of Ref.~\onlinecite{saul2014} and may be related to the enhanced hybridization between V $3d$ and Te $5p$ states. Te $5p$ orbitals contribute 6.5\,\% of states in the vicinity of the Fermi level, which is comparable to 9.7\,\% contributed by O $2p$ (see also Fig.~\ref{fig:dos}). A somewhat similar microscopic scenario has been reported for CdVO$_3$~\cite{tsirlin2011}, where $5s$ orbitals of Cd admix to the V $3d$ states and trigger ferromagnetic exchange couplings that give rise to the overall ferromagnetic long-range order, a very rare case among V$^{4+}$ oxides.

The couplings beyond $J_1$ are long-range. Their mechanism is usually understood as \mbox{V--O$\ldots$O--V} superexchange controlled by the V--O$\ldots$O angles defining the linearity of the superexchange pathway, and by the O$\ldots$O distance. Indeed, the larger values of $J_2$ and $J_{b2}$ can be ascribed to the shortest O$\ldots$O distances of $2.82-2.84$\,\r A and 2.87\,\r A, respectively. In contrast, the longer O$\ldots$O distance of 3.31\,\r A disfavors the interchain coupling $J_{b1}$, even though its V--V distance is 0.55\,\r A shorter than those of $J_2$ and $J_{b2}$. Finally, the weakly FM nature of $J_a$, $J_{a1}$, and $J_{a2}$ is typical for interactions in the direction perpendicular to basal planes of VO$_5$ pyramids, as in Pb$_2$V$_3$O$_9$~\cite{tsirlin2011b} and Zn$_2$VO(PO$_4)_2$~\cite{yusuf2010}. Here, no suitable V--O$\ldots$O--V pathway for an efficient $xy\rightarrow xy$ hopping can be formed, hence the AFM contribution is very small. On the other hand, the V--V distance is short enough to induce non-zero $xy\rightarrow yz$ and $xy\rightarrow xz$ hoppings resulting in weakly FM superexchange.

\begin{figure}
\includegraphics[width=\columnwidth]{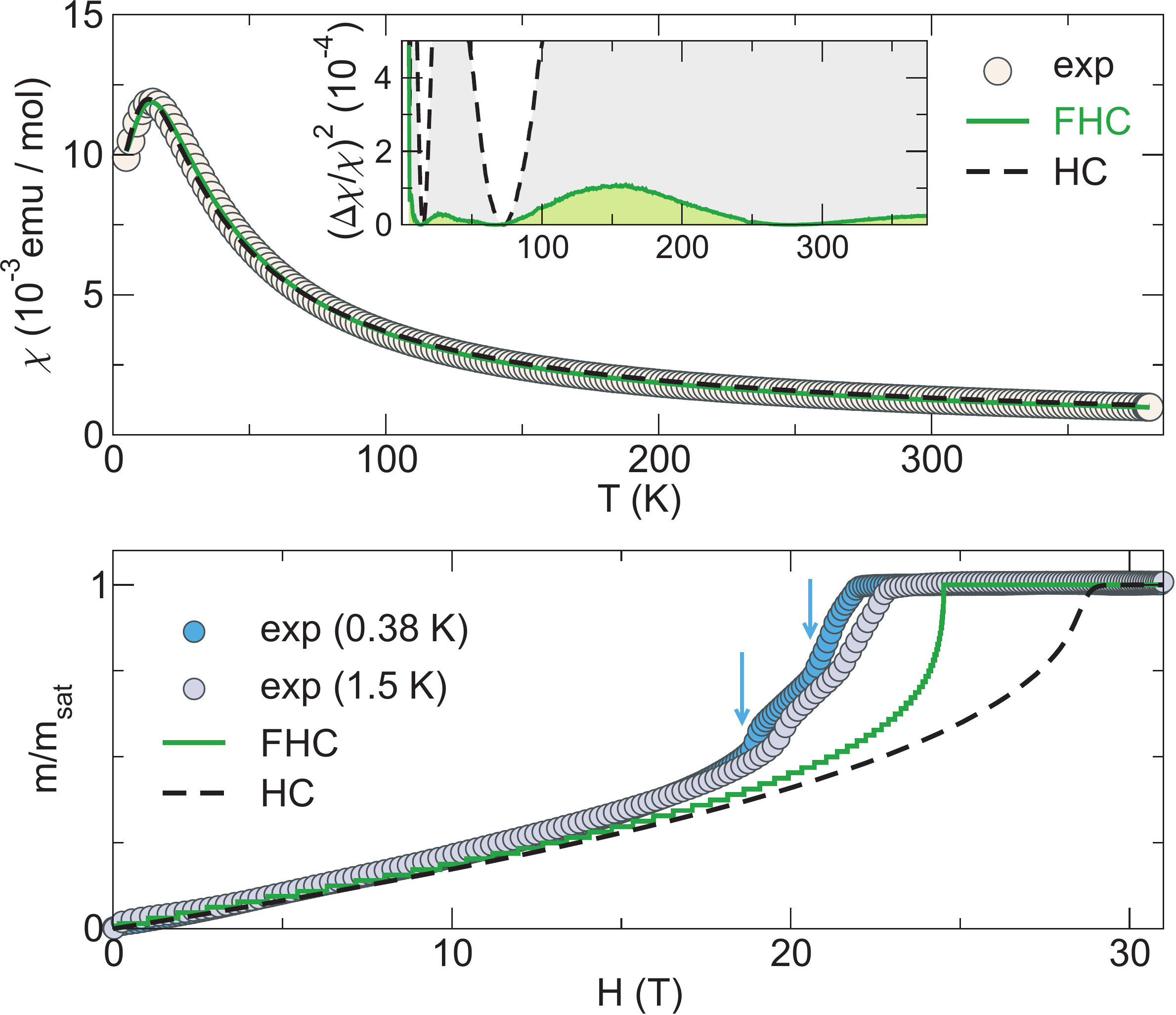}
\caption{\label{fig:fits}
(Color online) Top: experimental magnetic susceptibility fitted with the uniform Heisenberg chain (HC) and frustrated Heisenberg chain (FHC) models. The inset shows the difference between the experimental and simulated curves, as defined in the text. Bottom: experimental magnetization curve compared to the model predictions with no adjustable parameters.
}
\end{figure}

\subsubsection{Comparison to the experiment}
In $\beta$-TeVO$_4$, frustrated interactions $J_1$ and $J_2$ along the spin chains manifest themselves already in thermodynamic properties. In Fig.~\ref{fig:fits}, we compare both temperature dependence of the magnetic susceptibility and field dependence of the magnetization with two models: i) uniform Heisenberg chain (HC) proposed in Ref.~\onlinecite{savina2011}; and ii) $J_1-J_2$ frustrated Heisenberg chain (FHC) supported by our calculations. Both models provide good fits of the magnetic susceptibility resulting in $J=20.2$\,K, $g=2.09$ for the HC and $J_2=-J_1=26.4$\,K, $g=2.00$ for the FHC. However, a closer examination of the difference between the simulated and experimental curves $\left(\Delta\chi=(\chi_{\rm calc}-\chi_{\rm exp})^2/\chi_{\rm exp}^2\right)$ reveals better agreement for the FHC model (Fig.~\ref{fig:fits}, top inset). This model also provides a better description of the high-field magnetization curve, although the saturation field is slightly overestimated. Finally, the fitted $g$-value for the FHC model is in good agreement with $g=2.01$ measured by electron spin resonance~\cite{pregelj2015}, while the fitted $g$-value for the HC model would be too high. 

\begin{table}[!t]
\caption{\label{tab:dm}
DM couplings in $\beta$-TeVO$_4$. For the notation of the V$^{4+}$ sites $1-4$, see Figures~\ref{fig:structure} and~\ref{fig:dm}.
}
\begin{ruledtabular}
\begin{tabular}{c@{=}c@{\hspace{2em}}c@{=}c}
  $\Dv_1^{(12)}$  & $(d_{1x},d_{1y},d_{1z})$     & $\Dv_2^{(11')}$ & $(d_{2x},d_{2y},d_{2z})$    \\
	$\Dv_1^{(21')}$ & $(-d_{1x},-d_{1y},-d_{1z})$  & $\Dv_2^{(22')}$ & $(d_{2x},-d_{2y},d_{2z})$   \\
	$\Dv_1^{(34)}$  & $(d_{1x},d_{1y},d_{1z})$     & $\Dv_2^{(33')}$ & $(-d_{2x},-d_{2y},-d_{2z})$ \\
	$\Dv_1^{(34')}$ & $(-d_{1x},-d_{1y},-d_{1z})$  & $\Dv_2^{(44')}$ & $(-d_{2x},d_{2y},-d_{2z})$  \\
\end{tabular}
\end{ruledtabular}
\end{table}
\begin{figure}
\includegraphics{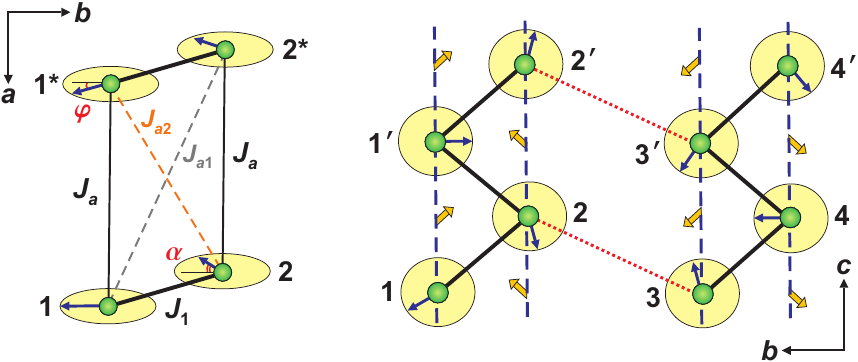}
\caption{\label{fig:dm}
(Color online) Formation of the helical magnetic structure in $\beta$-TeVO$_4$. Left panel: the competition between the helical structure along the chain and the FM interchain couplings ($J_a$, $J_{a1}$, $J_{a2}$), resulting in a phase shift $\varphi$ between the neighboring helices along the $a$ directions. Right panel: same direction of rotation (counter-clock-wise in this case) in the chains 1--2 and 3--4 is imposed by the AFM interchain coupling $J_{b2}$. It is, however, incompatible with the DM vectors $\Dv_2$ (shown by orange arrows) that change sign from ($d_{2x},\pm d_{2y},d_{2z})$ to $(-d_{2x},\mp d_{2y},-d_{2z})$ upon going from the chain 1--2 to the chain 3--4. Note that the $bc$ plane is arbitrarily chosen as the helix plane for the sake of better visualization.
}
\end{figure}
We can now rationalize the incommensurate magnetic order that has been observed in $\beta$-TeVO$_4$ experimentally~\cite{pregelj2015}. The propagation vector $\kv=(-0.208,0,0.423)$ below $T_{\rm N3}$ implies the pitch angle $\alpha=76.1^{\circ}$ for the helix propagating along the $c$ direction. This is in very good agreement with $J_2/J_1=-1$ that yields the classical pitch angle $\alpha_{\rm cl}=\arccos(-J_1/4J_2)=75.5^{\circ}$, whereas quantum corrections to the pitch angle are small in this range of $J_2/J_1$~\cite{zinke2009}. 

The leading interchain coupling along the $b$ direction, $J_{b2}$, is compatible with the helical order and leads to the same spin arrangement within every second chain, so that, e.g., both atoms~3 in Fig.~\ref{fig:structure} (middle) feature parallel spins resulting in $k_y=0$. On the other hand, the FM interchain couplings are incompatible with the helical order (Fig.~\ref{fig:dm}, left). The couplings $J_1$, $J_a$, and $J_{a1}$ ($J_{a2}$) build triangles, where the non-collinear order of the two spins coupled by $J_1$ cannot be combined with the FM order imposed by $J_a$ and $J_{a1}$ ($J_{a2}$). This frustration is alleviated by introducing a non-collinear spin arrangement on $J_a$ and $J_{a1}$ ($J_{a2}$) too. By a classical energy minimization for the magnetic model defined by the exchange couplings from Table~\ref{tab:exchanges}~\footnote{Here, we fixed the angle between spins 1 and 2 to the pitch angle of the helix $\alpha$. We checked that the simultaneous variation of $\varphi$ and $\alpha$ leads to a nearly indistinguishable result, because the value of $\alpha$ is fixed by a much larger energy scale of $J_2\simeq -J_1\simeq 25$\,K.}, we find that the spins on atoms 1 and $1^*$ should be turned by an angle
\begin{equation}
  \varphi=\text{arctan}\left[\frac{(J_{a2}-J_{a1})\sin\alpha}{(J_{a2}+J_{a1})\cos\alpha+J_a}\right]\simeq -53.9^{\circ}
\end{equation}	
corresponding to $k_x=\varphi/360=-0.150$, which is somewhat lower than $k_x=-0.208$ in the experiment, but shows the correct sign and explains the opposite directions of rotation along $a$ and $c$. 

We find that the incommensurability along the $a$ direction is controlled by a rather subtle difference between the diagonal interactions $J_{a1}$ and $J_{a2}$ that correspond to the bond vectors $(1,\Delta y,\frac12)$ and $(1,\Delta y,-\frac12)$, respectively, where $\Delta y$ is the difference between the $y$-coordinates of two neighboring V atoms along the chain. The interaction $J_{a1}$ is more FM than $J_{a2}$. Therefore, the angle $\varphi$ has to counteract the rotation introduced by $\alpha$ in order to bring spins 1 and 2* connected by $J_{a1}$ closer to the parallel configuration.

\subsubsection{Magnetic anisotropy}
In the following, we extend our microscopic analysis to the anisotropy parameters entering Eq.~\eqref{eq:ham}. Technically, these parameters could be evaluated in the same spirit of perturbation theory as in Eq.~\eqref{eq:kk}, following Refs.~\onlinecite{mazurenko2008,vasiliev2013}. However, the evaluation of magnetic anisotropy requires that hoppings to all four unoccupied $d$-orbitals, including $d_{x^2-y^2}$ and $d_{3z^2-r^2}$, are defined, which is not the case in $\beta$-TeVO$_4$, where the strong hybridization between Te $5p$ and V $3d$ orbitals prevents unambiguous modeling of the higher-lying $d$-bands. Therefore, we restrict ourselves to a qualitative analysis that provides useful insight into the role of magnetic anisotropy in $\beta$-TeVO$_4$.

The DM terms are allowed by symmetry for both intrachain couplings. Symmetry relations between different components of the $\Dv_1$ and $\Dv_2$ on individual lattice bonds are summarized in Table~\ref{tab:dm}. Helical ground state implies that same rotations occur between every two contiguous spins in a chain (Fig.~\ref{fig:dm}). Since all three components of $\mathbf D_1$ change sign within the chain, the nearest-neighbor DM couplings do not gain any energy from the helical order. The same argument shows that $d_{2y}$ cannot stabilize the helical order, because it changes sign between $\mathbf D_2^{(11)}$ and $\mathbf D_2^{(22)}$. On the other hand, $d_{2x}$ and $d_{2z}$ gain energy from the helical order and should thus stabilize the helix in the plane perpendicular to $(d_{2x},0,d_{2z})$ within the $1-2$ chain. However, the relevant DM components in the neighboring $3-4$ chains are $(-d_{2x},0,-d_{2z})$ implying that the helix with the opposite sense of rotation will be stabilized. On the other hand, the AFM interchain coupling $J_{b2}$ imposes same sense of rotation in the helices $1-2$ and $3-4$ (Fig.~\ref{fig:dm}, right). 

The sign change of $\Dv_2$ between the two contiguous chains implies that the helix plane in $\beta$-TeVO$_4$ cannot be uniquely defined when the DM anisotropy is considered. This striking observation should be parallel to the following. First, the helical order is stabilized only below $T_3$ upon a first-order phase transition that entails an abrupt shrinkage of the unit cell along the $b$- and $c$-directions implying a structural effect that probably alleviates this frustration. Second, neutron-scattering data indicate two different helix planes for the chains $1-2$ and $3-4$, respectively~\cite{zaharko2014}. This observation is hard to reconcile with the monoclinic crystal structure of $\beta$-TeVO$_4$, where these two chains are crystallographically equivalent (related by an inversion symmetry), and thus a structural distortion must be involved.

The incompatibility of the helical order with the DM anisotropy in $\beta$-TeVO$_4$ is also a plausible reason behind the delayed formation of the helical phase upon cooling in zero field. Other $J_1-J_2$ frustrated-chain compounds reported so far~\cite{schaepers2013} undergo a direct (and second-order) transition from the paramagnetic phase to the helically-ordered phase in zero field. In $\beta$-TeVO$_4$, however, the paramagnetic and helically-ordered phases are separated by the SDW phase (between $T_{\rm N1}$ and $T_{\rm N2}$) and by the long-period stripe order (between $T_{\rm N2}$ and $T_{\rm N3}$), see Fig.~\ref{fig:diagram-c}. We can also conclude that the helical order is largely destabilized in the magnetic field applied along the $c$-direction (compare the two panels of Fig.~\ref{fig:diagram-a}), which should then be the common direction of the two helices, i.e., the direction that is most vulnerable to the application of the magnetic field.

\section{Discussion and Summary}
\label{sec:summary}
$\beta$-TeVO$_4$ is a structurally perfect material prototype of the $J_1-J_2$ frustrated spin chain model. Our data rule out any structural distortions preceding magnetic transitions in this compound, and clearly exclude any change in the orbital state of V$^{4+}$ upon cooling. On the other hand, magnetic anisotropy triggered by the low crystallographic symmetry introduces a very complex behavior, especially in low magnetic fields. By comparing temperature-field phase diagrams obtained for different field directions, we conclude that the transitions at $T_{\rm N2}$ and $T_{\rm N3}$ are strongly direction-dependent and should be influenced or even triggered by the presence of magnetic anisotropy. The field evolution of $T_{\rm N1}$ is, at first glance, reminiscent of a conventional long-range AFM ordering and reveals no appreciable anisotropy. Finally, the high-field phase emerging above 18\,T turns out to be weakly dependent on the field direction and thus nearly isotropic.

The emergence of the high-field phase for different field directions indicates its relation to the physics of the isotropic $J_1-J_2$ spin chain. This phase can be ascribed to the multipolar order or nematic state envisaged in recent theoretical studies~\cite{sudan2009,zhitomirsky2010,sato2013}. The multipolar order should be robust with respect to the interchain couplings at $J_2/J_1\simeq -1$~\cite{nishimoto2015}, which renders $\beta$-TeVO$_4$ a good model material for studying high-field physics of the frustrated $J_1-J_2$ spin chain. Further investigation of this high-field phase is highly desirable. 

In lower fields, we observed that the helical phase is destabilized by the magnetic field $H\|c$ and gives way to a larger region of the long-period stripe order, while for other field directions the stripe phase shrinks to a small pocket visible in low magnetic fields only. Both helical and spin-density-wave phases are incommensurate along both $a$ and $c$~\cite{pregelj2015}. While the incommensurability along the $c$-direction is a natural result of the intrachain frustration, the incommensurability along $a$ can be understood as a competition of ferromagnetic interchain couplings with the helical (or spin-density-wave) order within the chain.

Regarding magnetic anisotropy, $\beta$-TeVO$_4$ is different from any other $J_1-J_2$ frustrated-spin-chain compound reported so far. Materials like CuGeO$_3$ and NaCu$_2$O$_2$ lack DM couplings completely, because inversion centers are found in the middle of both nearest-neighbor and next-nearest-neighbor bonds. In LiCuVO$_4$ and linarite, inversion symmetry forbids $\Dv_2$, whereas $\Dv_1$ may be non-zero, although its exact magnitude is still unknown. Finally, $\beta$-TeVO$_4$ has inversion centers between the chains only. Therefore, all DM couplings are non-zero, and the symmetry of $\Dv_2$ is compatible with the helical order, thus providing additional stabilization energy for each helix, but impeding the order between the helices. This frustration of anisotropic exchange couplings may be the crux of the $\beta$-TeVO$_4$ physics in low magnetic fields that awaits further investigation with direct methods, such as neutron scattering in applied magnetic field.

\acknowledgments
We are grateful to Peter Lemmens for initiating this work. We also acknowledge fruitful discussions with Andr\'es Sa\'ul, Myron B. Salamon, and Johannes Richter, and the provision of the beamtime by the ESRF. The National High Magnetic Field Laboratory Pulsed-Field Facility is supported by the National Science Foundation (NSF), the US Department of Energy (DOE), and the State of Florida through NSF Cooperative Grant DMR-1157490. AT was supported by the Federal Ministry for Education and Research via the Sofja Kovalevskaya Award of Alexander von Humboldt Foundation. The work in Tallinn was supported by the Estonian Research Council grants MTT77, PUT733, PUT210 and IUT23-7.

%\bibliography{beta-TeVO4}

\end{document}